\documentclass[aps,prx,twocolumn,groupedaddress,showpacs,longbibliography]{revtex4-2}
\usepackage{graphicx}
\usepackage{amsmath}
\usepackage{amsfonts}
\usepackage{amssymb}
\usepackage[normalem]{ulem}
\usepackage{xcolor}
\usepackage{soul}
\graphicspath{{./Figures/}}
\usepackage{float}
\usepackage[colorlinks=true, citecolor=black, linkcolor=black, urlcolor=black]{hyperref}

\def\bra#1{{\left\langle#1\right\vert}}

\def\ket#1{{\left\vert#1\right\rangle}}
\def\abs#1{{\left|#1\right|}}

\begin{document}

\title{Applications of Spin-Dependent Generalized Squeezing in\\Hybrid Spin-Oscillator Quantum Processors}
\author{Kai Shinbrough}
\email{kai.shinbrough@physics.ox.ac.uk}
\author{Donovan J. Webb}
\author{Iver R. \O vergaard}
\author{Oana B\u{a}z\u{a}van}
\author{Sebastian Saner}
\author{Christopher J. Ballance}
\author{Raghavendra Srinivas}
\email{raghavendra.srinivas@physics.ox.ac.uk}

\affiliation{Department of Physics, University of Oxford, Parks Road, Oxford OX1 3PU, United Kingdom}

\date{\today}

\begin{abstract}
Generalized squeezing interactions are foundational to quantum optics, and have recently come under experimental control in hybrid spin-oscillator quantum processors [O. B{\u{a}}z{\u{a}}van, \textit{et al.}, Nat. Phys. \textbf{22}, 757 (2026); S. Saner, \textit{et al.}, Phys. Rev. X \textbf{16}, 021049 (2026)]. These interactions open the door for new applications in the processing of discrete- and continuous-variable quantum information, four of which we propose and investigate in this work: geometric phase gates mediated by spin-dependent generalized squeezing acting on two spins and a common oscillator; genuine $N$-body spin interactions mediated by individually addressed spin-dependent generalized squeezing; oscillator thermometry via spin readout; and the preparation of high-fidelity quantum states of the oscillator via spin-dependent generalized squeezing and mid-circuit measurement. A unifying feature of these applications is the geometric phase induced by generalized squeezing interactions, which is nonlinear in the Fock occupation of the oscillator and the interaction order of the generalized squeezing. This work provides a foundation for fast, high-fidelity discrete- and continuous-variable quantum computation and sensing in the hybrid spin–oscillator platform via generalized squeezing.
\end{abstract}

\maketitle

\section{Introduction}

Hybrid spin–oscillator systems---such as trapped ions~\cite{wineland1998experimental,leibfried2003quantum}, superconducting qubits \cite{blais2004cavity, clarke2008superconducting,devoret2013superconducting, blais2021circuit}, and neutral atoms \cite{kaufman2012cooling,kaufman2021quantum,lienhard2025generation}---are among the leading platforms for applications in quantum computing, simulation, and sensing. Hybrid systems of this type allow for engineered coupling between spin degrees of freedom and the degrees of freedom of a bosonic oscillator mode, most commonly through spin-dependent interactions that are linear in the oscillator position and momentum, such as the spin-dependent force (SDF)~\cite{leibfried2003quantum,haljan2005spin}. These interactions can be used to couple weakly-interacting spin states via a shared oscillator, for example in photon-mediated coupling in superconducting qubits \cite{majer2007coupling,sillanpaa2007coherent,van2013photon,blais2021circuit} and neutral atoms coupled to optical cavities \cite{welte2017cavity,welte2018photon}, and phonon-mediated coupling in trapped-ion systems \cite{sorensen1999quantum,leibfried2003quantum,wineland1998experimental}. These linear interactions have been used to perform some of the highest-fidelity two-qubit entangling gates demonstrated to date \cite{clarke2021high,srinivas2021high,hughes2025trapped,ransford202698} and play a key role in the operation of the world’s most accurate optical atomic clocks~\cite{schmidt2005spectroscopy, marshall2025high}. More recently, interactions that instead use the spin to mediate interactions in the oscillator have also been explored theoretically~\cite{sutherland2021universal,lin2026preparing} and demonstrated experimentally~\cite{buazuavan2026squeezing, saner2026generating}, and hybrid protocols that make use of both the discrete variable of the spin and the continuous variable of the oscillator are an emerging area of research~\cite{liu2026hybrid,crane2024hybrid,bazavan2024synthetic,saner2025real,araz2025hybrid,nobakht2026noise}.

Looking beyond linear interactions, nonlinear spin–oscillator coupling offers the prospect of qualitatively new functionality \cite{katz2022n,rojkov2026stabilization,shapira2023robust,bond2025optimal}. Even in the absence of spin dependence, second-order squeezing interactions have led to a range of applications in quantum-enhanced sensing \cite{aasi2013enhanced,casacio2021quantum,burd2019quantum,burd2024experimental} and in the amplification of entangling operations \cite{burd2021quantum,ge2019trapped,ge2019stroboscopic}. Nonlinear spin-oscillator coupling beyond second-order squeezing---so-called spin-dependent generalized squeezing---is of fundamental importance in quantum optics \cite{braunstein1987generalized,ashhab2025properties} and represents the most general form of this resource. The spin-dependent generalized squeezing interactions we consider here are of the form:
\begin{equation}
    \hat{H}(t) = \frac{\hbar\Omega}{2} \hat{J}_{\theta,\phi} \left(\hat{a}^k e^{i\delta t} + \hat{a}^{\dagger k}e^{-i\delta t}\right),
    \label{eq_Hgensq}
\end{equation}
\noindent where $\Omega$ and $\delta$ are the Rabi frequency and detuning relative to the oscillator frequency, respectively, controlled by the amplitude, phase, and frequency of fields generating the interaction \cite{sutherland2021universal,buazuavan2026squeezing}. The collective spin operator $\hat{J}_{\theta,\phi} = \sum_i\hat{\sigma}_{\theta,\phi}^{(i)}$ conditions the interaction, where $\hat{\sigma}_{\theta,\phi}^{(i)} = \hat{\sigma}_x^{(i)}\sin\theta\cos\phi + \hat{\sigma}_y^{(i)} \sin\theta\sin\phi + \hat{\sigma}^{(i)}_z\cos\theta$ is a linear combination of Pauli operators for each spin $i$, and the oscillator subsystem is parameterized by bosonic creation and annihilation operators $\hat{a}^{\dagger}$ and $\hat{a}$, respectively, and integer interaction order $k\geq1$.

These spin-dependent generalized squeezing interactions have recently been demonstrated experimentally in the trapped-ion platform, enabling the generation of squeezed ($k=2$), trisqueezed ($k=3$), and quadsqueezed ($k=4$) states \cite{buazuavan2026squeezing}, as well as their superpositions \cite{saner2026generating}. In this article, we investigate a selection of complementary applications of spin-dependent generalized squeezing, with the aim of illustrating its potential as a versatile tool in hybrid quantum systems. A central theme of the applications we explore is the geometric phase generated by these interactions. While the set of applications considered here is not exhaustive, this work serves as a starting point for exploring the broader landscape of functionality enabled by spin-dependent generalized squeezing.

We consider four applications of spin-dependent generalized squeezing in two broad categories: (i) applications that leverage the geometric phase acquired by closed-loop trajectories of the oscillator to affect operations on the spin subsystem (Sec.~\ref{sec_gates} and \ref{sec_NBody}), and (ii) applications that leverage the same geometric phase in combination with projective spin readout to probe and modify the oscillator subsystem (Sec. \ref{sec_thermometry} and \ref{sec_BSP}). Here we consider the hybrid spin-oscillator system formed by atomic ions trapped in a one-dimensional harmonic potential \cite{ozeri2011trapped}, however these applications are readily generalizable to any hybrid system with access to nonlinear spin-oscillator coupling \cite{chang2020observation,eriksson2024universal,lienhard2025generation}. In Sec. \ref{sec_gates}, we consider two-qubit entangling gates mediated directly by generalized squeezing interactions, generalizing the standard geometric phase gate mediated by linear SDFs. In Sec. \ref{sec_NBody}, we consider individually addressed generalized squeezing interactions on $N$ ions coupled to a common oscillator, which we use to generate genuine $N$-body interactions and $N$-qubit entangling gates that are continuous in time. In Sec. \ref{sec_thermometry}, we consider generalized-squeezing--mediated thermometry of a single-mode oscillator and the precision-time tradeoff in the estimation of Fock and average thermal occupation as a function of interaction order. In Sec. \ref{sec_BSP}, we consider squeezing-mediated preparation of target oscillator states via mid-circuit spin measurement in the limit of strong and weak nonlinear spin-oscillator coupling, deriving an analytic expression for the normalized Fock state transfer function after each projective spin measurement. In Sec.~\ref{sec_conc}, we conclude and address the role of these applications in the hybrid spin-oscillator platform.

In what follows, we compute spin and oscillator dynamics in a variety of contexts under the evolution of the Hamiltonian $\hat{H}(t)$ in Eq.~\eqref{eq_Hgensq}. As $\hat{H}(t)$ does not commute with itself at all times for non-zero $\delta$, we consider either full numeric integration of the time-dependent Schr\"{o}dinger equation to evaluate the coupled system dynamics \cite{kramer2018quantumoptics}, or analytic calculation of the Magnus expansion of the unitary time evolution operator  \cite{magnus1954exponential} 
\begin{equation}\label{eq_Magnus}
\hat{U}(t) = \exp\left[ \Phi_1(t) + \Phi_2(t) + \Phi_3(t) + ... \right], 
\end{equation}
\noindent where the first-order term 
\begin{equation}
\begin{split}
    \Phi_1(t) 
    &= -\frac{i}{\hbar}\int_0^t dt_1\,\hat{H}(t_1) \\
    &= -\frac{\Omega}{2\delta}\hat{J}_{\theta,\phi}
       \left[
           \hat{a}^k \left(e^{i\delta t}-1\right)
           - \hat{a}^{\dagger k}\left(e^{-i\delta t}-1\right)
       \right]
\end{split}
\end{equation}
\noindent represents periodic spin-dependent generalized squeezing and anti-squeezing, and the second-order term 
\begin{equation}\label{eq_Magnus2nd}
\begin{split}
    \Phi_2(t)
    &= -\frac{1}{2\hbar^2}
       \int_0^t dt_1
       \int_0^{t_1} dt_2\,
       [\hat{H}(t_1),\hat{H}(t_2)] \\
    &= -\frac{i\Omega^2}{4\delta^2}
       \hat{J}_{\theta,\phi}^2
       \left[\hat{a}^k,\hat{a}^{\dagger k}\right]
       \left[\delta t-\sin(\delta t)\right]
\end{split}
\end{equation}
\noindent represents dynamic and geometric phase accumulation on the polarized spin subspace (the subspace with non-zero $\hat{J}_{\theta,\phi}$ eigenvalues). The third- and higher-order terms in this expansion can be made arbitrarily small with appropriate control over Rabi frequency and detuning.

\begin{figure*}[t]
        \centering
	\includegraphics[width=\textwidth]{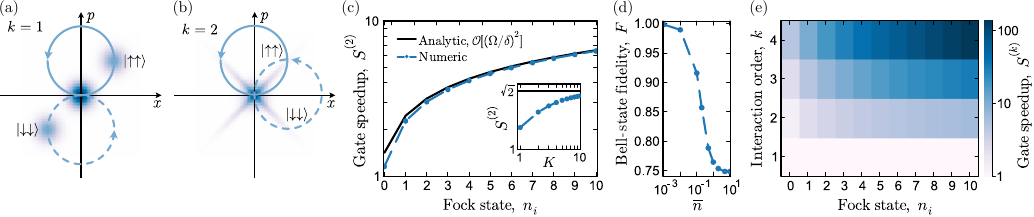}
    \vspace{-1.25em}
	\caption{Squeezing-mediated geometric phase gates. Wigner function trajectories for one gate duration are shown for spin-dependent (a) displacements ($k=1$), and (b) squeezing ($k=2$); Wigner functions are shown at $t_g^{(k)}/4$, with the amplitude and phase of the time-dependent displacement and squeezing interactions shown for $\ket{\uparrow\uparrow}$ (solid) and $\ket{\downarrow\downarrow}$ (dashed) spin states under $\hat{J}_z$ spin-conditioning. Spin states $\ket{\uparrow\downarrow}$ and $\ket{\downarrow\uparrow}$ remain at the origin for the duration of the gate. (c) Gate speedup for $k=2$ and $K=1$ as a function of initial Fock state occupation, $n_i$, assuming dynamics truncated at second order in the Magnus expansion (solid) and under full numeric integration of the underlying Hamiltonian (dashed, markers). Inset: convergence of analytic and numeric speedup for $n_i=0$ with number of phase-space loops per gate, $K$. (d) Bell-state fidelity for a $k=2$ interaction for thermal initial states of varying mean phonon number, $\bar{n}$. (e) Analytic gate speedup $S^{(k)}$ as a function of $k$ and $n_i$. Note logarithmic color scale.
 }
	\label{fig_gates}
\end{figure*}

\section{Squeezing-mediated geometric phase gates}\label{sec_gates}

For $k=1$, the Hamiltonian in Eq.~\eqref{eq_Hgensq} reproduces the well-known geometric phase gate in a two-spin, one-oscillator system \cite{sorensen1999quantum,sorensen2000entanglement, milburn2000ion, leibfried2003experimental}. In this case, the Magnus expansion of $\hat{U}(t)$ truncates after second order, and gate detunings of the form $\delta = 2\Omega\sqrt{K}$ (for integer number of phase space loops $K>0$) lead to the preparation of maximally entangled Bell states on the spin subsystem after the gate duration $t_g=2\pi K/\delta=\pi\sqrt{K}/\Omega$. As we show, these geometric phase gates generalize to arbitrary interaction order $k$, with a Fock-state-dependent gate speedup. From this point onward, we explicitly note the interaction order with a superscript in the generalized gate duration $t_g^{(k)}$ and Rabi frequency $\Omega^{(k)}$, and we define the generalized gate speedup $S^{(k)}=t_g^{(1)}/t_g^{(k)}$. Importantly, the ratio of interaction strength $\Omega^{(k)}/\Omega^{(1)}$ depends on the experimental implementation of the generalized squeezing interaction, and can be either linear \cite{sutherland2021universal,buazuavan2026squeezing} or nonlinear \cite{wineland1998experimental,katz2022n} in the Lamb-Dicke parameter. In what follows, we report two-qubit gate speedups for squeezing-mediated geometric phase gates in units of this ratio, where the breakeven point in gate speed relative to the $k=1$ geometric phase gate for a particular experimental implementation occurs when $S^{(k)} = \Omega^{(1)}/\Omega^{(k)}$. While a practical gate speedup depends on the experimental implementation, understanding the dynamics of these two-qubit gates and the geometric phase generated serves as a foundation for building up more sophisticated applications.

For $k\!\!\!=\!\!\!2$, an entangling gate mediated by spin-dependent squeezing, evaluating the terms of the Magnus expansion in Eq.~\eqref{eq_Magnus} to second order leads to an analytic gate duration of $t_g^{(2)} = \pi\sqrt{K}/[\Omega^{(2)}\sqrt{2(2n_i+1)}]$ and a gate speedup of $S^{(2)} = \left(\Omega^{(2)}/\Omega^{(1)}\right)\sqrt{2(2n_i+1)}$, for initial Fock state $n_i$. A comparison of the Wigner function trajectories for $k=1$ and $k=2$ geometric phase gates is provided in Fig.~\ref{fig_gates}(a) and (b), respectively [for further information, see Appendix A]. In the displacement-mediated ($k=1$) geometric phase gate, under $\hat{J}_z = \hat{\sigma}_z^{(1)}\otimes \mathbb{I}_s + \mathbb{I}_s \otimes \hat{\sigma}_z^{(2)}$ spin conditioning and a $\ket{++}\otimes\ket{n_i=0}$ initial state (where $\mathbb{I}_s$ is the identity operator acting on each spin and $\ket{n_i=0}$ is the vacuum state of the oscillator), the displacement of the $\ket{\uparrow\uparrow}$ (solid) and $\ket{\downarrow\downarrow}$ (dashed) spin components are $\pi$ out of phase, arising from the amplitude and phase of the generated $\alpha(t)$ in the displacement operator $\hat{D}[{\alpha}(t)] = \exp[\alpha(t) \hat{a}^\dagger - \alpha^*(t)\hat{a}]$ \cite{ozeri2011trapped}. In contrast, in the squeezing-mediated ($k=2$) gate the squeezing of the $\ket{\uparrow\uparrow}$ (solid) and $\ket{\downarrow\downarrow}$ (dashed) spin components are $\pi/2$ out of phase, arising from the amplitude and phase of the generated $\zeta(t)$ in the squeeze operator $\hat{s}[\zeta(t)] = \exp\{[\zeta(t)\hat{a}^{\dagger 2} - \zeta^*(t)\hat{a}^2]/2\}$ \cite{walls2025quantum}. Despite the fact that the centroid of the Wigner function is stationary at the origin for the duration of the squeezing-mediated gate, the non-stationary spin components nevertheless acquire a finite geometric phase from the regions of the Wigner function periodically squeezed away from the origin (for an analytic derivation of this effect based on infinitesimal Wigner function components, see Appendix B). 

In Fig.~\ref{fig_gates}(c), we compare the $\mathcal{O}[(\Omega/\delta)^2]$ analytic approximation of the squeezing-mediated gate speedup provided above to the empirical speedup resulting from full numeric integration of the time-dependent Schr\"odinger equation for $K=1$. The odd-order terms in Eq.~\eqref{eq_Magnus} beyond $\Phi_1(t)$ lead to residual periodic squeezing and anti-squeezing, and the even-order terms beyond $\Phi_2(t)$ lead to residual dynamic and geometric phase, resulting in a reduced numerical speedup for small $n_i$, and for small $K$, as shown in the inset. In both the numeric and analytic treatments, the Bell state fidelity $F = |\langle \phi | \psi \rangle|^2$ for $|\phi\rangle = (\ket{++}+i\ket{--})/\sqrt{2}$ remains unity for each Fock state occupation and number of loops, within a numerical accuracy of $\Delta F\sim10^{-3}$. As shown in the inset, the small difference in speedup between analytic and numeric calculation (for $n_i=0$, where the difference is the largest) is reduced with increasing number of phase space loops, $K$, consistent with this difference arising from the higher-order terms in Eq.~\eqref{eq_Magnus}:
As the terms of order $m$ in Eq.~\eqref{eq_Magnus} scale as $\mathcal{O}[(\Omega/\delta)^m]$, increasing $K$ truncates the magnitude of these terms relative to the desired $\mathcal{O}[(\Omega/\delta)]$ and $\mathcal{O}[(\Omega/\delta)^2]$ interactions. 

We consider the effect of thermal initial states on the $k=2$ squeezing-mediated entangling gate in Fig.~\ref{fig_gates}(d), optimizing over $\delta$ and $t_g$ for each average thermal occupation $\bar{n}$. In contrast to the $k=1$ entangling gate, which is robust to initial thermal occupation \cite{sorensen1999quantum,sorensen2000entanglement}, the $k=2$ entangling gate requires thermal occupations of $\bar{n}<10^{-2}$ to achieve Bell state fidelities $F>0.99$. These thermal states are at the limit of state-of-the-art sideband cooling and Fock state preparation \cite{rasmusson2021optimized,de2025modular}, and in Sec.~\ref{sec_thermometry} and \ref{sec_BSP} we discuss the complementary applications of high-fidelity thermometry and Fock state preparation enabled by spin-conditioned generalized squeezing, which can be used in sequence with squeezing-mediated geometric phase gates.

The scaling of the gate speedup $S^{(k)}=t_g^{(1)}/t_g^{(k)}$, where $t_g^{(k)}$ is the order-$k$ two-qubit entangling gate duration, is shown in Fig.~\ref{fig_gates}(e) for interaction order $k\leq4$ and initial Fock state occupation $n_i\leq10$. In the general case, the scaling of $S^{(k)}$ is superexponential in $k$ and polynomial in $n_i$ (see Appendix C).

\section{N-body spin interactions}\label{sec_NBody}

Thus far, we have considered pairwise entangling interactions between two spins. While these two-body interactions are sufficient for universal quantum computation when combined with individually addressed single-qubit rotations, the ability to directly engineer many-body interactions can unlock a broader range of applications in quantum simulation and sensing, and can reduce experimental overhead in certain quantum algorithms and error correcting codes \cite{locher2025multiqubit,zhang2025collective}. Despite this potential, methods for generating $N$-body interactions remain comparatively underdeveloped relative to two-qubit gates. Even in trapped-ion systems---among the most mature quantum computing platforms---experimental demonstrations of many-body interactions have only recently been achieved~\cite{katz2023demonstration}. In this section, we explore a novel approach to generate $N$-body interactions based on individually addressed spin-dependent generalized squeezing.

We assume a single oscillator mode with non-zero coupling to each spin, where any non-uniform spin-oscillator coupling across ions is absorbed into the Rabi frequency $\Omega_j$ for each individually addressed interaction. For notational simplicity, we use $\hat{\sigma}_j$ to represent the general Pauli operator $\hat{\sigma}_{\theta,\phi}^{(j)}$, where it is implied that $\theta$ and $\phi$ may differ for each spin $j$. We consider individually addressed $N$-body Hamiltonians of the form
\begin{equation}
    \hat{H}(t) = \sum_{j=1}^N \frac{\hbar\Omega_j}{2}\hat{\sigma}_j \left(\hat{a}^{k_j} e^{i\delta_j t} + \hat{a}^{\dagger k_j} e^{-i\delta_j t}\right),\label{eq_Hindvaddr}
\end{equation}
\noindent and two strategies for assignment of the interaction orders $k_j$ and detunings $\delta_j$. 

\textit{Assignment strategy (1)}---Resonant interaction between generalized squeezing of order $k_1=N-1$ on ion 1 and displacements on ions 2 through $N$ (i.e., $k_j=1$ for $j=2,...,N$). In this case, the desired resonant interaction leads to the constraint $\delta_1 - \sum_{j=2}^N \delta_j = 0$, and in order to suppress all spurious interactions involving fewer than $N$ ions we require all sum and difference terms involving fewer than $N$ detunings to be non-zero. These constraints are equivalent to the requirement of energy conservation for only the desired $N$-body interaction~\cite{sutherland2021universal}. The base case of this assignment strategy results in $\delta_1=(2^{N-1}-1)\Delta$, and $\delta_j=2^{N-j}\Delta$ for $j=2,...,N$ for a non-zero global detuning $\Delta$. 

\textit{Assignment strategy (2)}---Resonant interaction between $k_j=2$ squeezing on the first $N-2$ ions ($j=1,...,N-2$) and $k_i=1$ displacement on the last two ions ($i=N-1,N$). By inspection of the higher-order commutator structure in the Magnus expansion of Eq.~\eqref{eq_Magnus}, this assignment strategy leads to a non-zero energy-conserving $N$th order commutator for $\sum_{j=1}^N\delta_j(-1)^{j-1}=0$. Again we impose the constraint that sum and difference terms involving fewer than $N$ detunings should be non-zero in order to suppress spurious lower-order interactions involving fewer than $N$ ions. In this case, we consider the assignment strategy $\delta_j=2^{2(N-j)}\Delta$ for $j=2,...,N$, and $\delta_1 = \sum_{j=2}^N\delta_j(-1)^{j}$ in order to fulfill the $N$-body resonance condition. Although we leave a rigorous proof of spurious resonance avoidance for this strategy to future work, we confirm numerically that this assignment strategy is resonant only for the desired $N$-body interaction up to $N=100$. There may exist alternative detuning assignment strategies in this case that further minimize $\max(\delta_1,...,\delta_N)$, reducing the likelihood of spurious interaction with neighboring motional modes. 

Both assignment strategies \textit{(1)} and \textit{(2)} lead to an $N$th order term in the Magnus expansion of Eq.~\eqref{eq_Magnus} corresponding to an effective Hamiltonian of the form
\begin{equation}
    \hat{H}_\textrm{eff} = C_N(\Delta) \prod_{j=1}^N \Omega_j \hat{\sigma}_j,\label{eq_Heff3body}
\end{equation}
\noindent where the constant of proportionality $C_N(\Delta)$ varies depending on the assignment strategy and number of ions, and can be calculated for arbitrary order using standard non-commuting algebra \cite{de2012simplification}. The first $N-1$ terms in the Magnus expansion lead to either a global phase or, by construction, an off-resonant term that can be attenuated by increasing the global detuning, $\Delta$. All terms higher and lower in order than $N$ can be attenuated via pulse shaping of the interaction through the time-dependence of each $\Omega_j$ \cite{roos2008ion}. This procedure therefore leads to high-fidelity, genuine $N$-body interactions, independent of initial spin configuration, which are continuous in time and allow for significant reduction in circuit depth for many-body quantum computation and simulation \cite{nielsen2010quantum,barenco1995elementary,lloyd1996universal} (cf. related work in atomic systems \cite{luo2025realization,katz2022n,katz2023demonstration,katz2023programmable,roos2004control,bermudez2009competing,monz201114,shapira2020theory,arias2021high,andrade2022engineering}).

\begin{figure}[t]
    \centering
	\includegraphics[width=1\columnwidth]{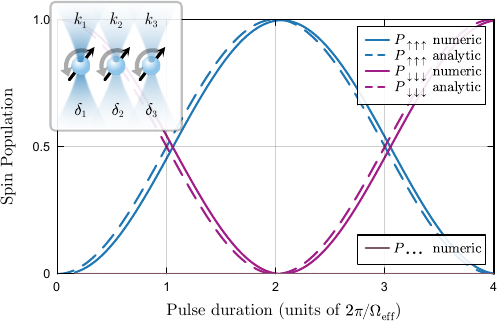}
    \vspace{-0.75em}
	\caption{Squeezing-mediated 3-body spin interactions. Spin population dynamics under the engineered 3-body interaction (solid) and analytic $\hat{\sigma}_x\hat{\sigma}_x\hat{\sigma}_x$ interaction (dashed). All 3-body spin populations are plotted; populations outside of the $P_{\uparrow\uparrow\uparrow}$ and $P_{\downarrow\downarrow\downarrow}$ subspace ($P_{\cdots}$) are smaller than the axis width. Inset: Schematic of individually addressed generalized squeezing interactions with squeezing ($k_1=2$) on ion 1, and displacements ($k_2=k_3=1$) on ions 2 and 3, with detunings $(\delta_1,\delta_2,\delta_3)=(3,2,1)\Delta$ for global detuning $\Delta$.} 
	\label{fig_3body}
\end{figure}

For the 3-body interaction shown in Fig.~\ref{fig_3body}, the two assignment strategies for $k_j$ and $\delta_j$ are identical, and lead to $(k_1,k_2,k_3)=(2,1,1)$ and the base case $(\delta_1,\delta_2,\delta_3)=(3,2,1)\Delta$. In this case, explicit calculation of the third-order Magnus expansion yields the proportionality constant $C_3(\Delta)=\hbar/(4\Delta^2)$ and an effective 3-body Rabi frequency $\Omega_\text{eff} = 2\Omega_1\Omega_2\Omega_3/\Delta^2$. In Fig.~\ref{fig_3body}, we show the spin population dynamics for 3 spins initialized in the state $\ket{\downarrow\downarrow\downarrow}$ with $\hat{\sigma}_x^{(j)}$ spin-conditioning for each ion, $j=1,2,3$. Here we employ a flat-top pulse shape for each $\Omega_j$ with a $\sin^2[\pi t/(2\tau)]$ ramp at the beginning and end of each pulse \cite{buazuavan2026squeezing,saner2026generating}, where we choose $\tau=1/\Omega_\text{eff}$. This short ramp duration leads to the offset between the full numeric simulation of Eq.~\eqref{eq_Hindvaddr} in the Schr\"{o}dinger equation [solid lines in Fig.~\ref{fig_3body}] and analytic time evolution according to $\hat{H}_\text{eff}(t) = (\hbar\Omega_\textrm{eff}/8)\hat{\sigma}_x^{(1)}\hat{\sigma}_x^{(2)}\hat{\sigma}_x^{(3)}$ [dashed lines, Fig.~\ref{fig_3body}]. The remaining population outside of the fully polarized subspace ($P_{\cdots}$) is plotted in Fig.~\ref{fig_3body}, but is smaller than the axis width for $\Delta=10 (2\Omega_1\Omega_2\Omega_3)^{1/3}$ and the chosen ramp duration. For simplicity we set $\Omega_1=\Omega_2=\Omega_3$, but we note that the effective total Rabi frequency of the interaction scales as the product in Eq.~\eqref{eq_Heff3body}, such that smaller Rabi frequencies for generalized squeezing may be compensated with stronger displacements on the remaining ions. For this combination of interaction parameters, we find numeric overlap infidelities $1- F_\textrm{GHZ}$ below $10^{-3}$, for $F_\textrm{GHZ} = \abs{\bra{\textrm{GHZ}}\psi\rangle}^2$, where $\ket{\psi}$ is the prepared spin state near $t=2\pi/\Omega_\textrm{eff}$ and $\ket{\textrm{GHZ}} =(\ket{\uparrow\uparrow\uparrow}-i\ket{\downarrow\downarrow\downarrow})/\sqrt{2}$ is a phase-rotated GHZ state \cite{greenberger1989going}. This infidelity is first-order insensitive to the Fock state occupation of the oscillator, which we verify by considering thermal initial states of the oscillator with $\bar{n}$ ranging from $10^{-3}$ to $10^{1}$; $1-F_\textrm{GHZ}$ is unchanged over this range, within numerical accuracy of $\Delta F_\textrm{GHZ}\sim10^{-3}$.

We also consider the use of multimode squeezing, displacements, and beamsplitting, which we find relax the constraints on $\delta_j$. For example, in the 3-body case, two-mode squeezing (or beamsplitting) on modes $a$ and $b$ on ion 1, displacement on mode $a$ on ion 2, and displacement on mode $b$ on ion 3 leads to the same effective $N$-body interaction generated via the methods above for a single mode, without the sum and difference constraints on detunings $\delta_2$ and $\delta_3$ as operators on mode $a$ and $b$ always commute. In general, the use of multimode interactions lifts the constraints on the detunings for those interactions that act on different modes, and therefore allows for smaller detunings, weaker spurious interactions with neighboring modes, and lesser optical power.

For 4-body interactions, as discussed in Appendix~D, these two assignments are non-identical, and there can be a significant difference in experimental complexity between $(k_1,k_2,k_3,k_4)=(3,1,1,1)$ and $(k_1,k_2,k_3,k_4)=(2,2,1,1)$, for assignment strategy \textit{(1)} and \textit{(2)}, respectively. In the general case, assignment \textit{(2)} allows for the construction of genuine, arbitrary-order $N$-body interactions with only $k=2$ squeezing as a resource, whereas assignment \textit{(1)} allows for the same $N$-body interactions with only one individually addressed generalized squeezing interaction, albeit at interaction order that increases linearly as $k_1=N-1$. As a point of comparison, the CNOT-ladder decomposition of the 3-body interaction above consists of 4 two-qubit gates and 7 single-qubit gates, while the same decomposition of the 4-body interaction consists of 6 two-qubit gates and 9 single-qubit gates; in the general case, a single $N$-body $X$-type interaction is equivalent to a circuit of $2(N-1)$ two-qubit gates and $2N+1$ single-qubit gates \cite{nielsen2010quantum}.

While we consider 3- and 4-body (in Appendix~D) $\hat{\sigma}_x$-type interactions here, the assignment strategies we develop represent a general protocol for engineering many-body interactions for any number of spins with any spin-conditioning.

\section{Squeezing-mediated thermometry}\label{sec_thermometry}

\begin{figure}[t]
    \centering
	\includegraphics[width=1.0\columnwidth]{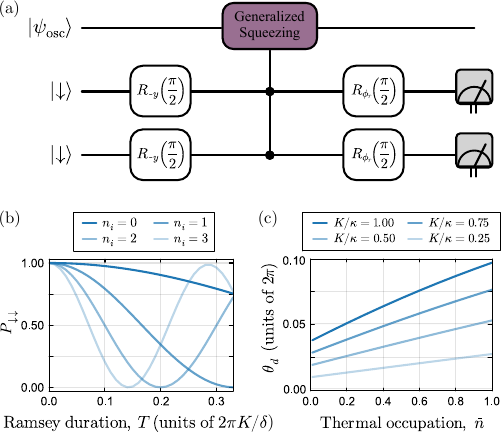}
	\caption{Squeezing-mediated thermometry. 
    (a) Circuit diagram depicting a spin-conditioned generalized squeezing interaction inside a Ramsey sequence on the spin subsystem. (b) Resulting Fock-state ($n_i$) dependent Ramsey fringes measured on the spin subsystem with readout phase $\phi_r=\pi/2$. (c) Differential phase accumulation for a thermal initial state of the oscillator for varying interaction strength parametrized by $\kappa$ (see text). 
    }
	\label{fig_thermometry}
\end{figure}

In Sec. \ref{sec_gates} and \ref{sec_NBody}, the geometric phase arising from generalized spin-dependent squeezing was engineered in order to generate effective spin-spin interactions. In this section and in Sec. \ref{sec_BSP}, we invert this engineering procedure in order to generate applications involving the bosonic mode mediated by the spin subsystem. Here we describe a simple Ramsey sequence by which the Fock-state-dependent geometric phase arising from spin-dependent generalized squeezing with $k\geq2$ can be used for bosonic mode thermometry, leaving extensions of this scheme to circuits of larger depth and comparison with other methods of bosonic thermometry (see, e.g., Refs.~\cite{rasmusson2021optimized,mallweger2024motional,mallweger2023single,li2024thermometry,bonetto2025numerical}) to future work. Bosonic thermometry, and bosonic mode characterization more generally, is crucial for any protocol based on spin-oscillator coupling; many errors in two-qubit entangling gates, for instance, are sensitive to the initial temperature of the mediating bosonic mode \cite{sutherland2022one,pino2021demonstration,ransford202698}, as are bosonic qubit encodings \cite{gottesman2001encoding,michael2016new,grimsmo2020quantum,gutman2025squeezed} and spin-boson quantum simulation protocols \cite{saner2025real,bazavan2024synthetic,rainaldi2026trigonometric,chalermpusitarak2025programmable,mcgarry2026programmable,navickas2025experimental,olaya2025simulating}. 

The circuit diagram for this scheme is shown in Fig.~\ref{fig_thermometry}(a), where the spin-dependent generalized squeezing interaction is applied between two global $\pi/2$ rotations on the spin subsystem. We consider the following example case with maximal contrast, wherein the first $\pi$/2 rotation, $R_{-y}(\pi/2)$, prepares the spin state $\ket{++}$ before a $\hat{J}_z$ spin-conditioned nonlinear interaction is applied for a Ramsey duration $T$ and a subsequent $\pi$/2 rotation is applied on the spin subsystem with readout phase $\phi_r$. The generalization to different initial $\pi/2$ pulse phases and spin-conditioning of the nonlinear interaction is immediate. For an oscillator with initial Fock state occupation $n_i$, we find the probability of measuring the $\ket{\downarrow\downarrow}$ spin state after discrete interaction times $T=2\pi K/\delta$ for arbitrary detuning $\delta^2=\kappa\Omega^2$ to be
\begin{equation}\label{eq_Pdowndown}
    P_{\downarrow\downarrow} = \frac{1}{8}\left(3 - \cos2 \phi_r + 4 \cos\theta_d \sin\phi_r\right),
\end{equation}
\noindent where $\theta_d$ is the time-dependent differential phase accumulated on the polarized spin subspace. Considering the first two terms of the Magnus expansion in Eq.~\eqref{eq_Magnus}, we derive the analytic result $\theta_d = 2\pi K \langle [\hat{a}^k,\hat{a}^{\dagger k}]\rangle/\kappa$, where the evaluation of the commutator expectation value for arbitrary $n_i$ and $k$ can be found in Appendix C. From this analytic result, we find the dimensionless ratio $K/\kappa$ to be the figure of merit determining the differential phase accumulation for a given initial temperature and interaction order, where $\kappa$ controls the amount of dynamic and geometric phase accumulated per phase-space loop, and $K$ controls the number of phase-space loops applied before readout. In Fig.~\ref{fig_thermometry}(b), we plot $P_{\downarrow\downarrow}$ for the $k=2$ interaction order, for initial Fock state occupations $n_i\in\{0,1,2,3\}$ and a readout phase of $\phi_r=\pi/2$, with $K/\kappa=0.25$. In Fig.~\ref{fig_thermometry}(c) we plot the differential phase $\theta_d$ measured through full simulation of the time-dependent Schr\"{o}dinger equation for thermal initial states of the oscillator, which exhibits the linear precision-time tradeoff expected for $k=2$ from the analytic treatment of the second order Magnus expansion (this tradeoff is quadratic in $\bar{n}$ for $k=3$, cubic for $k=4$, etc.). For both coherent and mixed states of the oscillator, the Fock occupation amplitudes can be extracted via the Fourier transform of the time-domain Ramsey scans evaluated at frequencies $\omega_{n} = \Omega\langle [\hat{a}^k,\hat{a}^{\dagger k}]\rangle/\sqrt{\kappa}$ where the commutator expectation value is evaluated at Fock occupation $n$.

In the case of projection onto a bright spin state, fluorescent readout induces random displacements onto the oscillator in integer multiples of the fluorescent photon momentum. If desired, these random displacements can be pre-empted via state transfer of the oscillator to a protected mode \cite{hou2024coherent}. In this case, and in the case of dark state projection, the Ramsey sequence of Fig.~\ref{fig_thermometry}(a) nevertheless induces backaction on the oscillator due to interference between spin components that acquire geometric phase and spin components that do not. This effect can be leveraged for the preparation of oscillator states, as discussed in Sec.~\ref{sec_BSP}, or for bosonic quantum error correction codes where the spin readout can be engineered to project the oscillator onto the logical state subspace \cite{leghtas2013hardware,mirrahimi2014dynamically}.

\section{Bosonic state preparation}\label{sec_BSP}

The backaction of the projective spin measurement in the circuit of Fig.~\ref{fig_thermometry}(a) can be formulated as a Fock state map acting on the oscillator subspace useful for preparing a broad class of bosonic states. Spin measurements that result in projection onto the polarized subspace map pure states of the oscillator, $\ket{\psi} = \sum_{n=0}^\infty c_n \ket{n}$ (in the Fock basis $\{\ket{n}\}$), onto the family of pure states $\ket{\psi'} = (1/\mathcal{N})\sum_{n=0}^\infty t_n c_n \ket{n}$, where $\mathcal{N}$ is a normalization constant, $t_n = 1 + e^{-i\theta_d} \sin\phi_r$, and $\theta_d$ is, as before, the Fock-state-dependent geometric phase accumulated by the oscillator during the Ramsey sequence. In effect, $t_n$ represents a coherent and diagonal Fock state transfer function applied to the oscillator after each round of the circuit, which can be optimized in an oscillator state preparation sequence composed of multiple concatenated circuit rounds. Each round of this concatenated circuit possesses independent and tuneable readout and geometric phases, parametrized by $\phi_r$ and $k$, $\kappa$, and $K$, respectively. 

Here we consider two applications of this bosonic state preparation sequence: preparation of high-fidelity Fock states from a thermal oscillator distribution, and preparation of 4-component Schr\"{o}dinger cat states from a coherent displaced state of the oscillator. The preparation of additional non-trivial oscillator states from alternative initial distributions generalizes readily. 

\subsection{Fock state preparation}
The preparation of individual Fock states is a central capability in the control of quantum harmonic oscillators. Fock states are of fundamental interest due to their strongly nonclassical features, such as Wigner negativity and non-Gaussianity; they provide a useful benchmark for oscillator state engineering and high-fidelity bosonic control; and they serve as valuable resources for applications in quantum-enhanced sensing and the speedup of two-qubit gates, as discussed in Sec.~\ref{sec_gates}.

\begin{figure}[t]
    \centering
	\includegraphics[width=1.0\columnwidth]{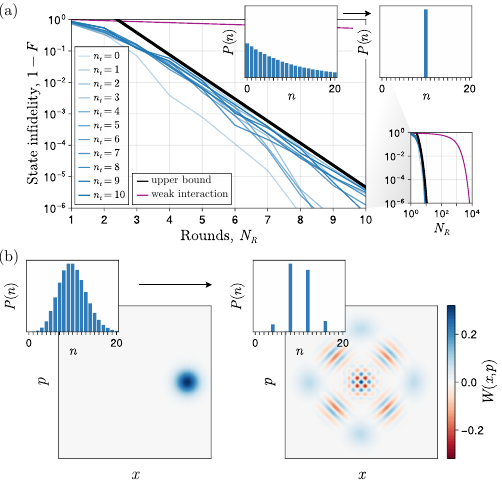}
	\caption{Squeezing-mediated bosonic state preparation. (a) Fock state preparation infidelity in the strong interaction protocol (see text) versus number of rounds of the bosonic state preparation sequence, $N_R$, for target Fock states $n_t=0$ to $10$, assuming an initial thermal oscillator state with $\bar{n}=10$ (phonon number distribution shown in the inset). An exponential fit to the state infidelity bound is provided (black), in addition to the infidelity for $n_t=0$ in the weak interaction protocol (magenta and lower right panel). (b) Wigner functions and phonon number distributions for a coherent state prepared into the symmetric 4-component cat state.
 }
	\label{fig_stateprep}
\end{figure}

For the preparation of high-fidelity Fock states, we consider two circuit optimization strategies we refer to as the weak and strong interaction protocols, respectively. In both cases, we generalize the treatment above to mixed states of the oscillator, and consider the transfer function acting on the phonon number distribution, $T_{\downarrow\downarrow}(n)\propto\abs{t_n}^2$, where after each round of the circuit we find a phonon number distribution of $P'(n) = T_{\downarrow\downarrow}(n)P(n)$ with respect to the pre-circuit distribution, $P(n)$. In the weak interaction protocol, each round of the preparation sequence is identical, where $\phi_r$, $k$, $\kappa$, and $K$ are chosen such that $T_{\downarrow\downarrow}(n)$ is maximized for the target Fock state and has a bandwidth comparable to the average thermal state phonon number $\bar{n}$, such that the next maximum of $T_{\downarrow\downarrow}(n)$ (which is periodic in $n$) occurs at a negligible thermal occupation. This protocol ensures asymptotically pure Fock state preparation at the cost of a large number of rounds and thus protocol time, as shown in the lower right panel of Fig.~\ref{fig_stateprep}(a) (magenta line) for ground state preparation from an initial thermal state with mean phonon number $\bar{n}=10$ and a Ramsey sequence with $\phi_r=\pi/2$, $k=2$, and $K/\kappa=2.5\times10^{-3}$. In the strong interaction protocol, by contrast, $\phi_r$, $k$, $\kappa$, and $K$ may differ each round, and are optimized jointly to maximize target Fock state preparation fidelity for a fixed number of rounds $N_R$. Here we train a genetic algorithm \cite{conn1997globally} to minimize state preparation infidelity with respect to the restricted set of parameters $\{\phi_r, k, \kappa\}^{N_R}$, assuming fixed number of phase space loops $K=1$. This restriction is trivial to relax in experiment, as any per-round target differential phase $\theta_d^T$ can be generated with an arbitrary number of loops by scaling the per-loop phase by $K$. We define the state infidelity $1-F=1-\textrm{Tr}[\mathbb{I}_s\otimes\ket{n_t}\bra{n_t}\rho]$ for the spin-oscillator state $\rho$, where $n_t$ is the target Fock occupation and $\mathbb{I}_s$ is the identity operator on the spin subspace, and we further restrict $2\leq k\leq 4$ and $\kappa\geq1$ to align the optimizer with experimental feasibility \cite{buazuavan2026squeezing,saner2026generating}. In Fig.~\ref{fig_stateprep}(a), we show the optimized preparation infidelity resulting from this algorithm for target Fock states $n_t=0$ to $10$ for an initial thermal occupation of $\bar{n}=10$ and $N_R$ ranging from 1 to 10. Generally, the lower occupation target Fock states have lower preparation infidelity, as the initial thermal state possesses a larger initial overlap with these target states, but all target Fock states considered here achieve $\mathcal{O}(10^{-5})$ infidelity or better after 10 rounds of the state preparation circuit. Qualitatively, the optimized set $\{\phi_r, k, \kappa\}^{N_R}$ found by the optimizer corresponds to an alternating strategy of broadband Fock state transfer functions with small $\theta_d$ followed by narrowband transfer functions with large $\theta_d$, where $\phi_r$ is modulated to keep the maximum of the transfer function centered on the target Fock state. The broadband transfer functions shift the center of the Fock state distribution function closer to the target, while the narrowband transfer functions remove population in Fock states neighboring the target. The optimal strong protocol requires a combination of these two effects, where the constraints on $\phi_r, k, \kappa$, and $N_R$ lead to non-trivial dynamics for a generic $n_t$. While the strategy found by the optimizer alternates between broadband and narrowband transfer functions, the transfer functions are diagonal and can be performed in arbitrary order with no change to the resulting state infidelity. For $N_R>5$, we provide an upper bound on the state preparation infidelity by fitting the maximum of $1-F$ across $n_t$ to an exponential distribution [shown in black in Fig.~\ref{fig_stateprep}(a)]; we find the infidelity improves exponentially with the number of rounds applied, bounded above by $1-F\leq0.2^{N_R}$.

Both the weak and strong interaction protocols provide a herald for successful state preparation via spin readout. Here we consider only the simple case of $N_R$ successful $\ket{\downarrow\downarrow}$ spin projections, each occurring with probability $P_{\downarrow\downarrow}$ given in Eq.~\eqref{eq_Pdowndown}, resulting in a cumulative success probability $P_{N_R} =\prod_{i=1}^{N_R} P_{\downarrow\downarrow}$. An adaptive strategy with real-time feedback from arbitrary spin projection is also possible, and may be the subject of future work. Narrowband Fock state transfer functions can also be used as a final `clean-up' operation on thermal states with small average phonon occupation (for example, prepared by sideband cooling and blue-sideband excitation) to significantly improve Fock state preparation fidelity with a small number of rounds and high success probability.

\subsection{Cat state preparation}

The preparation of Schr\"{o}dinger cat states of an oscillator is of importance to fundamental physics \cite{sanders1992entangled,monroe1996schrodinger,bild2023schrodinger}, quantum error correction \cite{ralph2003quantum,michael2016new,mirrahimi2014dynamically,rojkov2026stabilization,fong2025engineering}, and quantum sensing \cite{munro2002weak}. A key feature of cat states is their phonon number parity---the 2-component cat states $\ket{C^{\pm}_\alpha} = (\ket{\alpha}\pm\ket{-\alpha})/\mathcal{N}$ (with normalization constant $\mathcal{N}$) possess finite support on only the even and odd Fock states, respectively, for $\ket{C^{+}_\alpha}$ and $\ket{C^{-}_\alpha}$. The even parity 4-component cat states $\vert C^{(n\!\bmod\!4)}_\alpha\rangle = (\ket{C^\pm_\alpha}\pm\ket{C^\pm_{i\alpha}})/\mathcal{N}$ possess finite support only on Fock states with $n\bmod4=0$ or $2$, with the two logical basis states in the 4-component cat code occupying the $0\bmod4$ and $2\bmod4$ subspaces, respectively \cite{mirrahimi2014dynamically,michael2016new}. These states all share the same amplitude and phase envelope given by the parent coherent state $\ket{\alpha}$, and can thus be distilled from the parent state via filtration of the occupied Fock states into the appropriate subspace. In this work, we identify that these parity and amplitude restrictions are natively compatible with the periodic Fock state transfer functions achieved through spin-dependent generalized squeezing via the state preparation circuit described above. We note that many useful non-classical oscillator states possess similar parity and envelope restrictions (squeezed states and binomial states, among others). 

As an illustrative example of this technique, we consider the distillation of a 4-component cat state from a coherent state of the oscillator. As shown in Fig.~\ref{fig_stateprep}(b), we consider a parent coherent state before the state preparation sequence with mean phonon number $\abs{\alpha}^2=10$. After $N_R=5$ circuit rounds, we distill the $\vert C^{(0)}_\alpha\rangle = (\ket{C^+_\alpha}+\ket{C^+_{i\alpha}})/\mathcal{N}$ 4-component cat state with infidelity $1-F<10^{-2}$ and cumulative success probability of $P_{N_R}=0.152$ without active feedback. We show the initial and final numerically constructed Wigner functions in Fig.~\ref{fig_stateprep}(b), along with the phonon number distributions corresponding to the constructed states, the latter of which is well approximated by the product of the initial phonon number distribution and a filter function with $n\bmod4=0$. Qualitatively, the optimized Fock state transfer functions found by the genetic algorithm in this case again correspond to an alternating strategy, this time between transfer functions that null Fock components with $n\bmod2=1$ (and thus $n\bmod4=3$) and transfer functions that null $n\bmod4=1$ and $2$, leaving only components with appreciable amplitude at $n\bmod4=0$.

\section{Conclusion}\label{sec_conc}

Since their introduction, linear interactions that couple the spin degrees of freedom of hybrid systems to harmonic oscillator degrees of freedom have become a workhorse across trapped ions, superconducting circuits, neutral atoms, and related platforms, enabling a wide range of applications in quantum computing, simulation, and sensing. Extending this hybrid control beyond linear coupling, through out-of-Lamb-Dicke quantum control \cite{rojkov2026stabilization,kasri2026nonlinear,cheng2018nonlinear,de1996nonlinear}, higher-order sideband engineering \cite{katz2022n,katz2023demonstration,katz2023programmable}, amplitude- and phase-modulated spin-dependent forces \cite{matsos2024robust,millican2025engineering,matsos2025universal}, and non-commuting spin-dependent forces \cite{sutherland2021universal,rainaldi2026trigonometric,chalermpusitarak2025programmable,fong2025engineering,mcgarry2026programmable}, further expands the set of operations that can be realized. In this work, building on recently demonstrated spin-dependent generalized squeezing realized via non-commuting spin-dependent forces \cite{buazuavan2026squeezing,saner2026generating}, we have introduced four complementary applications of this new type of nonlinear spin-oscillator coupling: squeezing-mediated geometric phase gates, genuine $N$-body spin interactions, thermometry, and bosonic state preparation. These applications all stem from the underlying structure of the spin-dependent generalized squeezing Hamiltonian in Eq.~\eqref{eq_Hgensq} and its integrated commutators in the Magnus expansion of Eq.~\eqref{eq_Magnus}.  While in this work we consider two ions (or $N$ ions, as is the case in Sec.~\ref{sec_NBody}) coupled to a single oscillator mode, the results presented here are readily generalizable to systems of arbitrary spin and oscillator number.

Taken together, these examples illustrate the versatility of spin-dependent generalized squeezing as a resource for hybrid quantum information processing. The squeezing-mediated two-qubit gates of Sec.~\ref{sec_gates} combined with the high-fidelity Fock state preparation of Sec.~\ref{sec_BSP}, as well as the $N$-body spin interactions of Sec.~\ref{sec_NBody}, provide routes toward fast, high-fidelity discrete-variable quantum computation and simulation. By contrast, the squeezing-mediated thermometry and oscillator state preparation protocols of Sec.~\ref{sec_thermometry} and \ref{sec_BSP} contribute to the growing toolbox for continuous-variable quantum computation, simulation, and sensing. While each application considered here uses the hybrid spin-oscillator interaction to generate an effective operation primarily on either the spin or oscillator subsystem, the underlying dynamics intrinsically involve both. We anticipate that further applications will emerge in regimes where the spin and oscillator are used simultaneously as active quantum resources, for example in hybrid quantum simulations, bosonic error correction, and protocols that combine discrete- and continuous-variable information processing within a single device.

\section*{Acknowledgements}

We thank Gabriel Araneda, Emily Hirsch, Giorgio Canalella, Robert Tyler Sutherland, and Ana Maria Rey for helpful discussions. This work was supported by the US Army Research Office (W911NF-20-1-0038 and W911NF-25-1-0010) and the UK EPSRC Hub in Quantum Computing and Simulation (EP/T001062/1). CJB acknowledges support from a UKRI FL Fellowship. RS acknowledges funding from an EPSRC Fellowship EP/W028026/1.

\section*{Appendix A: Further information on $k=1,2$ Wigner function trajectories}

Full animations of the phase-space dynamics discussed in the main text are provided in the Supplemental Material as the video file $\texttt{gate\_wigner\_function\_dynamics.mp4}$. The animations show the time evolution of the Wigner function under the gates of interaction order (a) $k=1$, $\delta = 2\sqrt{K}\Omega$ and (b) $k=2$, $\delta \approx 1.15\times2\sqrt{K}\Omega$, with spin components $\ket{\uparrow\uparrow}$, $\ket{\uparrow\downarrow}$, $\ket{\downarrow\uparrow}$, and $\ket{\downarrow\downarrow}$ indicated in color for $\hat{J}_z$ spin conditioning and a $\ket{++}$ initial spin configuration.

\section*{Appendix B: Derivation of Geometric Phase Accumulation in the $(k=2)$ Squeezing-Mediated Geometric Phase Gate}

Consider the first order term in the Magnus expansion of the displacement-mediated geometric phase gate Hamiltonian [Eq.~\eqref{eq_Hgensq}, with $k=1$]:
\begin{align}
    \hat{U}_1(t) &= \exp\left\{-\frac{\Omega}{2\delta}\hat{J}_{\theta,\phi} \left[\hat{a} (e^{i\delta t} - 1) - \hat{a}^{\dagger}(e^{-i\delta t} - 1)\right]\right\},
\end{align}
\noindent which describes a displacement operation with $\alpha(t) = \Omega\langle \hat{J}_{\theta,\phi}\rangle(e^{-i\delta t} - 1)/(2\delta)$. This interaction affects a affine transformation of the $\hat{x}$ and $\hat{p}$ phase space coordinates
\begin{align}
    \begin{bmatrix} 
        \hat{x}\\  
        \hat{p} 
    \end{bmatrix} 
    \rightarrow 
    \hat{\textbf{D}}(\alpha)
    \begin{bmatrix} 
        \hat{x}\\  
        \hat{p} 
    \end{bmatrix} = 
    \begin{bmatrix} 
        \hat{x}\\  
        \hat{p} 
    \end{bmatrix} +
    \begin{bmatrix} 
        \sqrt{2}\,\text{Re}(\alpha)\\  
        \sqrt{2}\,\text{Im}(\alpha) 
    \end{bmatrix},
\end{align}
\noindent where
\begin{align}
    \sqrt{2}\,\text{Re}(\alpha) &= \sqrt{2}\Omega\langle \hat{J}_{\theta,\phi}\rangle[\cos(\delta t) - 1]/(2\delta)\\
    \sqrt{2}\,\text{Im}(\alpha) &= -\sqrt{2}\Omega\langle \hat{J}_{\theta,\phi}\rangle\sin(\delta t)/(2\delta),
\end{align}
\noindent where we assume real $\Omega$ without loss of generality.

This affine transformation describes cyclic motion of the phase space coordinates. To calculate the area, $A$, enclosed by one set of coordinates $(x_0,p_0)$ in one period, we consider the line integral over the closed loop $\Gamma$ (which is equivalent to the area enclosed via Green's theorem):
\begin{equation}
\begin{split}
    A &= \frac{1}{2} \oint_\Gamma (x\,dp - p\,dx)\\
    &= \frac{1}{2} \int_{t=0}^{2\pi/\delta} \left(x(t)\,\frac{dp}{dt} - p(t)\,\frac{dx}{dt}\right) dt\\
    &= -\frac{\Omega^2}{4\delta}\langle \hat{J}_{\theta,\phi}\rangle^2 \int_{t=0}^{2\pi/\delta} \left[\cos^2(\delta t) + \sin^2(\delta t)\right] dt\\
    &= \frac{\Omega^2}{4\delta^2}\langle \hat{J}_{\theta,\phi}\rangle^2 (2\pi),
\end{split}
\end{equation}
\noindent which agrees with the second order term in the Magnus expansion evaluated after a closed loop up to an overall sign, and represents a closed-loop geometric phase. As shown in Ref.~\cite{ozeri2011trapped}, the second order term in the Magnus expansion represents the total phase accumulation on the hybrid system, and can be decomposed into dynamical and geometric parts, where the dynamical phase of $2\Phi_2(t)$ and geometric phase of $-\Phi_2(t)$ sum to provide the total phase $\Phi_2(t)$. The area enclosed in this case is independent of the initial coordinates $(x_0,p_0)$, which makes the integral corresponding to the average loop closure over all coordinates weighted by the Wigner function trivial:
\begin{equation}
\begin{split}
    \bar{A}
    &= \frac{
        \int_{-\infty}^{\infty} dx_0
        \int_{-\infty}^{\infty} dp_0\,
        A(x_0,p_0)W(x_0,p_0)
    }{
        \int_{-\infty}^{\infty} dx_0
        \int_{-\infty}^{\infty} dp_0\,
        W(x_0,p_0)
    } \\
    &= A.
\end{split}
\end{equation}
Next, we consider the $k=2$ case using the same approach. The first order term in the Magnus expansion of the squeezing-mediated geometric phase gate Hamiltonian [Eq.~\eqref{eq_Hgensq}, with $k=2$] is:
\begin{align}
    \hat{U}_1(t) &= \exp\left\{-\frac{\Omega}{2\delta}\hat{J}_{\theta,\phi} \left[\hat{a}^2 (e^{i\delta t} - 1) - \hat{a}^{\dagger 2}(e^{-i\delta t} - 1)\right]\right\},
\end{align}
\noindent which describes a squeezing interaction with $\zeta(t) = \Omega\langle \hat{J}_{\theta,\phi}\rangle(e^{-i\delta t} - 1)/\delta$. This interaction affects a symplectic transformation of the $\hat{x}$ and $\hat{p}$ phase space coordinates
\begin{align}
    \begin{bmatrix} 
        \hat{x}\\  
        \hat{p} 
    \end{bmatrix} 
    \rightarrow 
    \hat{\textbf{S}}(\zeta)
    \begin{bmatrix} 
        \hat{x}\\  
        \hat{p} 
\end{bmatrix},
\end{align}
\noindent where
\begin{align}
    \hat{\textbf{S}}(\zeta) =
\begin{bmatrix}
\cosh r - \cos\theta \sinh r & -\sin\theta \sinh r \\
-\sin\theta \sinh r & \cosh r + \cos\theta \sinh r
\end{bmatrix},
\end{align}
\noindent and $\zeta=r e^{i\theta}$ \cite{walls2025quantum}. Assuming real $\Omega$ without loss of generality, this implies
\begin{align}
    r &= 2\Omega\langle \hat{J}_{\theta,\phi}\rangle\sin(\delta t/2)/\delta,\\
    \theta &= -(\delta t/2 + \pi/2).
\end{align}
This symplectic transformation again describes cyclic motion of the phase space coordinates. To calculate the area enclosed by one set of coordinates $(x_0,p_0)$ in one period, we again consider the line integral over the closed loop $\Gamma$:
\begin{equation}
\begin{split}
    A &= \frac{1}{2} \oint_\Gamma (x\,dp - p\,dx)\\
    &= \frac{1}{2} \int_{t=0}^{2\pi/\delta} \left(x(t)\,\frac{dp}{dt} - p(t)\,\frac{dx}{dt}\right) dt\\
    &= \frac{\pi}{4
    } \left[1 + 2 \frac{\Omega }{\delta }\langle \hat{J}_{\theta,\phi}\rangle \right.\\
    &\hspace{4em} \left.- I_0\left(4 \frac{\Omega }{\delta } \langle \hat{J}_{\theta,\phi}\rangle \right) - I_1\left(4 \frac{\Omega }{\delta } \langle \hat{J}_{\theta,\phi}\rangle \right) \right]x^2_0 \\
    &\hspace{1em} + \frac{\pi}{4}\left[1 - 2 \frac{\Omega}{\delta} \langle \hat{J}_{\theta,\phi}\rangle \right. \\
    &\hspace{4em}\left. - I_0\left(4 \frac{\Omega}{\delta} \langle \hat{J}_{\theta,\phi}\rangle \right) + I_1\left(4 \frac{\Omega}{\delta} \langle \hat{J}_{\theta,\phi}\rangle \right) \right] p^2_0,
\end{split}
\end{equation}
\noindent where $I_0(x)$ and $I_1(x)$ are modified Bessel functions of the first kind. 

In this case, $A(x_0,p_0)$ is dependent on the initial coordinates $(x_0,p_0)$, and $A(x_0,p_0)\rightarrow\infty$ as $x_0\rightarrow\infty$ or $p_0\rightarrow\infty$, which necessitates averaging over a finite scalar field (i.e., the Wigner function). In order to consider the average loop closure over all coordinates weighted by an arbitrary initial Wigner function, we consider the complete basis of Fock state Wigner functions:
\begin{align}
    W_n(x_0,p_0) = \frac{2}{\pi}(-1)^n \mathcal{L}_n\left[2(x_0^2+p_0^2)\right]e^{-(x_0^2+p_0^2)},
\end{align}
\noindent where $n$ is the Fock state occupation and $\mathcal{L}_n(x)$ is the Laguerre polynomial of degree $n$. Using this basis, we calculate the average loop closure
\begin{equation}
\begin{split}
    \bar{A}(n) &= \frac{\int_{-\infty}^{\infty}dx_0\int_{-\infty}^{\infty}dp_0\, A(x_0,p_0) W_n(x_0,p_0)}{\int_{-\infty}^{\infty}dx_0\int_{-\infty}^{\infty}dp_0\, W_n(x_0,p_0)}\\
    &= \frac{\pi}{8}(2n+1) \left[2 - 2I_0\left(4 \frac{\Omega }{\delta } \langle \hat{J}_{\theta,\phi}\rangle \right) \right]\\
    &\approx \frac{\Omega^2}{2\delta^2} \langle \hat{J}_{\theta,\phi}\rangle^2 (2n+1) (2\pi),
\end{split}
\end{equation}
\noindent where in the final line we have used the Taylor expansion $I_0(x)\approx1+x^2/4 +\mathcal{O}(x^4)$ for small $\Omega/\delta$ (valid for $\delta=2\sqrt{K}\Omega$). This result again agrees with the dynamic and geometric phase resulting from the second order term in the Magnus expansion evaluated after a closed loop.

\section*{Appendix C: Speedup scaling in generalized squeezing-mediated Geometric Phase Gates}

For arbitrary $k$ in Eq.~\eqref{eq_Hgensq}, the second term in the Magnus expansion of the unitary time evolution operator is given in Eq.~\eqref{eq_Magnus2nd}. For a maximally entangling gate, we set the geometric phase to $\pi/2$. At integer multiples of the motional period, $t=2\pi K/\delta$, we find the condition for a maximally entangling gate to be $\delta = 2\Omega^{(k)} \sqrt{K} \sqrt{\langle[\hat{a}^k,\hat{a}^{\dagger k}]\rangle}$, where the commutator expectation value is taken with respect to the initial state of the oscillator. In units of $\Omega^{(1)}/\Omega^{(k)}$, this corresponds to a gate speedup of 
\begin{align}
    S^{(k)} &= \sqrt{\langle[\hat{a}^k,\hat{a}^{\dagger k}]\rangle}\\
    \label{eq_speedup_analytic}&= \sqrt{(n_i+1)^{\overline{k}} - n_i^{\underline{k}}},
\end{align}
\noindent where $x^{\overline{k}}$ and $x^{\underline{k}}$ are the rising and falling factorial functions, respectively. For $n_i=0$, the scaling of the constant term in $S^{(k)}$ is
\begin{align}
    S^{(k)}\big\vert_{n_i=0} = \sqrt{k!},
\end{align}
\noindent which is super-exponential in $k$. For fixed $k$, the largest power of $n_i$ in $S^{(k)}$ is
\begin{align}
    S^{(k)} \sim\mathcal{O}[kn_i^{(k-1)/2}],
\end{align}
\noindent which is polynomial in $n_i$.

\newpage
\section*{Appendix D: Squeezing-mediated 4-body interactions}

\begin{figure}[t!]
    \centering
	\includegraphics[width=1\columnwidth]{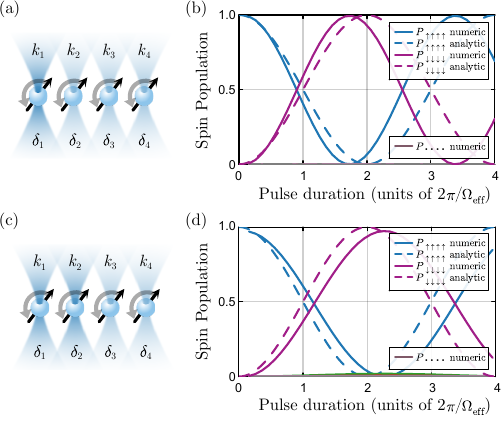}
	\caption{Squeezing-mediated 4-body spin interactions. Schematics of individually addressed generalized squeezing interactions with (a) trisqueezing ($k_1=3$) on ion 1, and displacements ($k_2=k_3=k_4=1$) on ions 2 through 4, and (c) squeezing ($k_1=k_2=2$) on ions 1 and 2, and displacements ($k_3=k_4=1$) on ions 3 and 4. (b), (d) Spin population dynamics under the engineered 4-body interactions (solid) and analytic $\hat{\sigma}_x\hat{\sigma}_x\hat{\sigma}_x\hat{\sigma}_x$ interaction (dashed). 
    }
	\label{fig_4body}
\end{figure}

In Fig.~\ref{fig_4body}, we show the numeric and analytic results of a 4-body $\hat{\sigma}_x^{(1)}\hat{\sigma}_x^{(2)}\hat{\sigma}_x^{(3)}\hat{\sigma}_x^{(4)}$-type spin interaction generated via spin-dependent generalized squeezing using both assignment strategy \textit{(1)} and \textit{(2)} discussed in the main text. In Fig.~\ref{fig_4body}(a) and (b), we consider strategy \textit{(1)}: individually addressed tri-squeezing ($k_1=3$) applied to ion 1, and displacements ($k_2=k_3=k_4=1$) applied to ions 2, 3, and 4, with $(\delta_1,\delta_2,\delta_3,\delta_4)=(13,8,4,1)\Delta$ and $\Delta=3(3\Omega_1\Omega_2\Omega_3\Omega_4/16)^{1/4}$, with a ramp duration of $1/\Omega_\textrm{eff}$, where $\Omega_\textrm{eff}=3\Omega_1\Omega_2\Omega_3\Omega_4/(16\Delta^3)$ [i.e., $C_4(\Delta)=3\hbar/(128\Delta^3)]$. In Fig.~\ref{fig_4body}(c) and (d), we consider strategy \textit{(2)}: individually addressed squeezing ($k_1=k_2=2$) applied to ions 1 and 2, and displacements ($k_3=k_4=1$) applied to ions 3 and 4, with $(\delta_1,\delta_2,\delta_3,\delta_4)=(13,16,4,1)\Delta$ and $\Delta=3(\Omega_1\Omega_2\Omega_3\Omega_4)^{1/4}$, with a ramp duration of $1/\Omega_\textrm{eff}$, where $\Omega_\textrm{eff}=413\Omega_1\Omega_2\Omega_3\Omega_4/(63648\Delta^3)$ [i.e., $C_4(\Delta)=413\hbar/(509184\Delta^3)]$. 
Demonstrating the limitations of assignment strategy \textit{(2)} for finite ramp duration and interaction strength, the spin dynamics in Fig.~\ref{fig_4body}(d) exhibit significant population error. 
A comprehensive study of the error budget for the $N$-body gates in both assignment strategies may be the subject of future work.

\newpage

\bibliography{bibliography}

\end{document}